\documentclass[12pt,twoside]{article}
\usepackage{amsmath, amsthm, amssymb, mathrsfs} 
\usepackage{graphics, graphicx}

\newcommand{\partdif}[2]{\ensuremath{ \frac{\partial #1}{\partial #2}}}

\begin{document}

\title{Extremum-Preserving Limiters for MUSCL and PPM}
\author{Michael Sekora \\ 
\textit{{\small Program in Applied and Computational Mathematics, Princeton University}} \\ 
\textit{{\small Princeton, NJ 08540, USA}} \\
Phillip Colella \\ 
\textit{{\small Applied Numerical Algorithms Group, Lawrence Berkeley National Laboratory,}} \\ 
\textit{{\small 1 Cyclotron Road, Berkeley, CA  94720, USA}}}
\date{25 March 2009}
\maketitle

\pagestyle{myheadings}
\markboth{M Sekora, P Colella}{Extremum-Preserving Limiters for MUSCL and PPM}


\noindent
Limiters are nonlinear hybridization techniques that are used to preserve positivity and monotonicity when numerically solving hyperbolic conservation laws. Unfortunately, the original methods suffer from the truncation-error being $1^{st}$ order accurate at all extrema despite the accuracy of the higher-order method \cite{borisBook:1976,vanLeer:1977,zalesak:1979,harten:1982}. To remedy this problem, higher-order extensions were proposed that relied on elaborate analytic and geometric constructions \cite{ENO,WENO,huynh,riderETAL:2007}. Since extremum-preserving limiters are applied only at extrema, additional computational cost is negligible. Therefore, extremum-preserving limiters ensure higher-order spatial accuracy while maintaining simplicity. This report presents higher-order limiting for $(i)$ computing van Leer slopes and $(ii)$ adjusting parabolic profiles. This limiting preserves monotonicity and accuracy at smooth extrema, maintains stability in the presence of discontinuities and under-resolved gradients, and is based on constraining the interpolated values at extrema (and only at extrema) by using nonlinear combinations of $2^{nd}$ derivatives. The van Leer limiting can be done separately and implemented in MUSCL (Monotone Upstream-centered Schemes for Conservation Laws) \cite{vanLeer:1977} or done in concert with the parabolic profile limiting and implemented in PPM (Piecewise Parabolic Method) \cite{colellaWoodward:1984,millerColella:2002}. The extremum-preserving limiters elegantly fit into any algorithm which uses conventional limiting techniques. Limiters are outlined for scalar advection and nonlinear systems of conservation laws. This report also discusses the $4^{th}$ order correction to the point-valued, cell-centered initial conditions that is necessary for implementing higher-order limiting. The material herein complements Colella and Sekora \cite{colellaSekora:2007}. Lastly, there is no guarantee that extremum-preserving limiters preserve positivity. To ensure this property, one should combine the limiting with FCT (Flux-Corrected Transport) \cite{zalesak:1979}.


\section{Algorithms for Scalar Advection}
\noindent
Consider the following scalar equation in one spatial dimension:
\begin{equation}
\partdif{a}{t} + \lambda \partdif{a}{x} = 0, ~~ \lambda = constant .
\end{equation}

\noindent
At time-step $n$, the average-valued, cell-centered quantity $a$ over a finite volume of length $h = \Delta x$ is:
\begin{equation}
a^n_i \approx \frac{1}{h} \int^{(i+\frac{1}{2})h}_{(i-\frac{1}{2})h} a(x,n \Delta t) dx .
\end{equation}

\noindent
MUSCL/PPM are conservative finite volume methods that are used to compute $a^{n+1}_i$:
\begin{equation}
a^{n+1}_i = a^n_i - \frac{\lambda \Delta t}{\Delta x} \left( a^{n+\frac{1}{2}}_{i+\frac{1}{2}} - a^{n+\frac{1}{2}}_{i-\frac{1}{2}} \right) ,
\end{equation}

\noindent
where $a^{n+\frac{1}{2}}_{i+\frac{1}{2}}$ is the average of a linear/parabolic interpolant over the interval swept out by the characteristics crossing the cell face at $(i+\frac{1}{2})h$ and is given by:
\begin{equation}
a^{n+\frac{1}{2}}_{i+\frac{1}{2}} = \left \{  \begin{array}{ll} 
a^{+}_{i+\frac{1}{2}} = \frac{1}{\sigma h} \int^{(i+\frac{1}{2})h}_{(i+\frac{1}{2}-\sigma)h} a^{I}_{i}(x) dx & \lambda \geq 0 \\
a^{-}_{i+\frac{1}{2}} = \frac{1}{\sigma h} \int^{(i+\frac{1}{2}+\sigma)h}_{(i+\frac{1}{2})h} a^{I}_{i+1}(x) dx & \lambda \leq 0
\end{array} \right . ,
\end{equation}

\noindent
where $\sigma = \frac{|\lambda| \Delta t}{\Delta x} \in [0,1]$ is the CFL number and $a^{I}_{i}(x)$ is the linear/parabolic interpolant, such that $x \in [(i-\frac{1}{2})h,(i+\frac{1}{2})h]$. 
\\

\noindent
There are three variations of Godunov-type methods for which limiters can be implemented:
\begin{enumerate}
\item van Leer Limiter in MUSCL
\item van Leer Limiter + Parabolic Profile Limiter in PPM
\item Parabolic Profile Limiter in PPM
\end{enumerate}
\noindent
Each of these algorithm variations are discussed below.

\subsection{MUSCL}
\begin{enumerate}
\item van Leer limit the differences $(\Delta a_i)$, giving $2^{nd}$ order results. Apply the corresponding boundary conditions.
\item Use the van Leer limited differences to compute $4^{th}$ order differences:
\begin{equation}
\Delta_{4} a_i = \frac{2}{3} \left( (a_{i+1} - \frac{1}{4}\Delta a_{i+1}) - (a_{i-1} + \frac{1}{4}\Delta a_{i-1}) \right) .
\end{equation}

\item Employ piecewise linear reconstruction by computing spatially extrapolated face-centered values at the low and high (left and right) edges of cells:
\begin{equation}
a_{i,\pm} = a_i + \frac{1}{2} \left( \pm 1 - \sigma \right) \Delta_{4} a_i .
\end{equation}

\end{enumerate}

\subsection{PPM}

\begin{enumerate}
\item van Leer limit the differences $(\Delta a_i)$, giving $2^{nd}$ order results. Apply the corresponding boundary conditions.
\item Employ either $4^{th}$ or $6^{th}$ order piecewise parabolic reconstruction:

\begin{itemize}
\item Method 1: use the van Leer limited differences and compute spatially extrapolated face-centered values at the low and high (left and right) edges of cells:
\begin{eqnarray}
{}^{4}a_{i,+} &=& \frac{1}{2}\left( a_{i+1} + a_{i} \right) - \frac{1}{6} \left( \Delta a_{i+1} - \Delta a_{i} \right) , \\
{}^{4}a_{i,-} &=& \frac{1}{2}\left( a_{i} + a_{i-1} \right) - \frac{1}{6} \left( \Delta a_{i} - \Delta a_{i-1} \right) , \\
{}^{6}a_{i,+} &=& {}^{4}a_{i,+} - \frac{1}{30} \left( 3(\Delta a_{i+1} - \Delta a_{i}) - (\Delta a_{i+2} - \Delta a_{i-1}) \right) , \\
{}^{6}a_{i,-} &=& {}^{4}a_{i,-} - \frac{1}{30} \left( 3(\Delta a_{i} - \Delta a_{i-1}) - (\Delta a_{i+1} - \Delta a_{i-2}) \right) , \\
\alpha_{\pm}  &=& a_{\pm} - a_i .
\end{eqnarray}

\item Method 2: employ either $4^{th}$ or $6^{th}$ order piecewise parabolic reconstruction without using the van Leer limited differences in Step 1:
\begin{eqnarray}
{}^{4}a_{i+\frac{1}{2}} &=& \frac{7}{12} \left( a_{i+1} + a_{i} \right) - \frac{1}{12} \left( a_{i+2} + a_{i-1} \right), \\
{}^{6}a_{i+\frac{1}{2}} &=& \frac{37}{60} \left( a_{i+1} + a_{i} \right) - \frac{8}{60} \left( a_{i+2} + a_{i-1} \right) + \frac{1}{60} \left( a_{i+3} + a_{i-2} \right) , \\
\alpha_{\pm} &=& a_{i \pm \frac{1}{2}} - a_{i} .
\end{eqnarray}

\noindent 
It is important to note that when $\Delta a_{i} = \Delta_{c} a_{i}$ (centered difference) the two formulations for the piecewise parabolic reconstruction are identical.
\end{itemize}

\item Limit the parabolic profile $(\alpha_{\pm})$.
\item Use the PPM predictor values to reconstruct the parabolic profile:
\begin{equation}
a_{\pm} = a_i + \alpha_{\pm} + \frac{\sigma}{2} (\pm(\alpha_{-} - \alpha_{+})) - (\alpha_{-}+\alpha_{+})(3-2\sigma) .
\end{equation}

\end{enumerate}

\subsection{Update Solution for MUSCL or PPM}

\begin{enumerate}
\item Compute fluxes:
\begin{equation}
F_{i+\frac{1}{2}} = \left \{  \begin{array}{ll} 
\lambda a_{+} ~~ \lambda \geq 0 , \\
\lambda a_{-} ~~ \lambda \leq 0 .
\end{array} \right . .
\end{equation}

\item Use the divergence of the fluxes to update the solution: 
\begin{eqnarray}
a^{n+1}_i &=& a^{n}_i - \frac{\Delta t}{\Delta x} \left( F_{i+\frac{1}{2}} - F_{i-\frac{1}{2}} \right) , \\
          &=& \left \{  \begin{array}{ll} 
a^{n}_i - \sigma \left( a^{+}_{i+\frac{1}{2}} - a^{+}_{i-\frac{1}{2}} \right) ~~ \lambda \geq 0 , \\
a^{n}_i - \sigma \left( a^{-}_{i+\frac{1}{2}} - a^{-}_{i-\frac{1}{2}} \right) ~~ \lambda \leq 0 .
\end{array} \right . .
\end{eqnarray}

\end{enumerate}

\subsection{Conventional Limiters}
\noindent
To complete the specification of the MUSCL and PPM schemes, one defines the conventional limiters used to constrain the interpolated profiles within each cell.

\subsubsection{Conventional van Leer Limiter}
\noindent
Given a sequence of average-valued, cell-centered quantities, the conventional van Leer limiter proceeds with the following steps \cite{vanLeer:1977,colellaWoodward:1984,millerColella:2002}:

\begin{enumerate}
\item Compute one-sided and centered differences:
\begin{eqnarray}
\Delta_{-} a_i &=& a_{i} - a_{i-1} , \\
\Delta_{c} a_i &=& \frac{1}{2} (a_{i+1} - a_{i-1}) , \\
\Delta_{+} a_i &=& a_{i+1} - a_{i} .
\end{eqnarray}

\item Apply the conventional van Leer limiter:
\begin{eqnarray}
\Delta_{\lim} a_i &=& 2 \min(|\Delta_{-} a_i|,|\Delta_{+} a_i|) \label{eq:vl} , \\
\mathcal{S}       &=& \textrm{sign}(\Delta_{c} a_i) , \\
\Delta a_{i}      &=& \left \{  \begin{array}{ll} 
\min(|\Delta_{c} a_i|,|\Delta_{\lim} a_i|) \mathcal{S} & \Delta_{-} a_i ~ \Delta_{+} a_i > 0 \\
0                                                      & \Delta_{-} a_i ~ \Delta_{+} a_i \leq 0
\end{array} \right . \label{eq:lv_ext} .
\end{eqnarray}

\end{enumerate}

\noindent
One significant defect of this method is the clipping of extremum when $\Delta_{-} a_i ~ \Delta_{+} a_i \leq 0$. This clipping sets $\Delta a_{i} \rightarrow 0$ as a precautionary measure for suppressing spurious oscillations. 
\\

\noindent
How one arrives at the formula for the conventional van Leer limiter can be understood by considering the following example. Assume that one is not at an extremum, $\frac{da}{dx} |_{ih} < 0$, and $|\Delta_{-} a_i|>|\Delta_{+} a_i|$. The value of $a((i+\frac{1}{2})h)$ at face-center $(i+\frac{1}{2})h$ is approximated by:
\begin{equation}
\begin{array}{ccccccc} 
a((i+\frac{1}{2})h) & = & a((i+1)h) & - & \frac{h}{2} \frac{da}{dx}|_{(i+1)h} & + & \cdots \\
a((i+\frac{1}{2})h) & = & a(ih)     & + & \frac{h}{2} \frac{da}{dx}|_{ih}     & + & \cdots 
\end{array}
\end{equation}

\noindent
Clearly, $a((i+\frac{1}{2})h) \geq a((i+1)h)$ and $a((i+\frac{1}{2})h) \approx a(ih) + \frac{h}{2} \frac{da}{dx}|_{ih}$. Therefore:
\begin{equation}
h \frac{da}{dx}|_{ih} \geq 2 (a((i+1)h) - a(ih)),
\end{equation}

\noindent
and one arrives at Eq. \ref{eq:vl}. However, at extrema $\frac{da}{dx} \rightarrow 0$, which leads one to Eq. \ref{eq:lv_ext}.

\subsubsection{Conventional Parabolic Profile Limiter}

Given the higher-order reconstruction of $a_{\pm}$ such that $\alpha_{\pm} = a_{\pm} - a_{i}$, the conventional parabolic profile limiter proceeds with the following steps \cite{colellaWoodward:1984,millerColella:2002}:

\begin{enumerate}
\item Adjust $\alpha_{\pm}$ according to the following cases:
\begin{equation}
\begin{array}{ccc}
\alpha_{\pm} \rightarrow 0  ~~~~~~     &  & \alpha_{+} \alpha_{-} \geq 0 , \\
\alpha_{+}   \rightarrow -2 \alpha_{-} &  & \alpha^{2}_{+} > 4 \alpha^{2}_{-} , \\
\alpha_{-}   \rightarrow -2 \alpha_{+} &  & \alpha^{2}_{-} > 4 \alpha^{2}_{+} .
\end{array}
\end{equation}

\item Reconstruct $a_{\pm}$ given the adjusted values for $\alpha_{\pm}$.
\end{enumerate}

\noindent
One significant defect of this method is the adjustments to $\alpha_{\pm}$ make the reconstruction a monotone profile as a precautionary measure for suppressing spurious oscillations. However, this constraint is more restrictive than is required to preserve monotonicity \cite{woodward:1986}.
\\

\noindent
The adjustments to $\alpha_{\pm}$ are derived from the interpolation polynomial described in the original Piecewise Parabolic Method:
\begin{eqnarray}
a_{i}          &=& a_{-} + \sigma (\delta_{\pm} a + a_{6} (1-\sigma)) , \\
\delta_{\pm} a &=& a_{+} - a_{-} = \alpha_{+} - \alpha_{-} , \\
a_{6}          &=& 6 \left( a_{i} - \frac{1}{2} (a_{+} + a_{-}) \right) = -3 (\alpha_{+} + \alpha_{-}) .
\end{eqnarray}

\noindent
where $\sigma \in [0,1]$ is the CFL number, which also corresponds to a dimensionless length scale. This scale is associated with each grid cell such that the left side of a cell is designated $\sigma = 0$ and the right side of a cell is designated $\sigma = 1$. Differentiating $a_{i}$ with respect to $\sigma$ gives:
\begin{equation}
\frac{d a_i}{d \sigma} = \delta_{\pm} a + a_{6} (1 - 2 \sigma) = (\alpha_{+} - \alpha_{-}) - (\alpha_{+} + \alpha_{-})(1 - 2 \sigma) .
\end{equation}

\noindent
By evaluating $d a_{i} / d \sigma$ at the left and right sides of the cell:
\begin{eqnarray}
\frac{d a_i}{d \sigma} \Big |^{-}_{\sigma \rightarrow 0} &=& \delta_{\pm} a + a_{6} = -2 (\alpha_{+} + 2 \alpha_{-}) , \\
\frac{d a_i}{d \sigma} \Big |^{+}_{\sigma \rightarrow 1} &=& \delta_{\pm} a - a_{6} =  2 (2 \alpha_{+} + \alpha_{-}) .
\end{eqnarray}

\noindent
Maximize $a_{i}$ with respect to $\sigma$ by setting the derivatives equal to zero and solving the resulting equations. One arrives at the following result:
\begin{equation}
\begin{array}{ccc}
\alpha_{+}   \rightarrow -2 \alpha_{-} &  & \sigma \rightarrow 0  , \\
\alpha_{-}   \rightarrow -2 \alpha_{+} &  & \sigma \rightarrow 1  .
\end{array}
\end{equation}


\section{Extremum-Preserving Limiters} 

\noindent
For smooth solutions away from extrema, the MUSCL scheme is $2^{nd}$ order accurate for linear advection whereas the PPM is $3^{rd}$ order accurate for linear advection and $4^{th}$ order accurate in the limit of vanishing CFL number. However, the monotonicity constraints at extrema reduce the truncation error to $O(h)$ even at smooth extrema. This reduction in the overall accuracy of the method also introduces a non-smooth component to the error. To eliminate this problem, one constructs a new limiting scheme at extrema.
\\

\noindent
The defect of the standard approach to limiting is most easily seen in the MUSCL limiter. Away from extrema, the magnitude of the slope is computed as the minimum of three undivided differences: the centered difference and twice the one-sided differences. In smooth regions away from extrema, the centered difference and the one-sided differences all approximate $h \frac{d a}{dx}$ such that the minimum is always defined by the centered difference. At discontinuities, one of the one-sided differences is typically much smaller than the other two differences. Therefore, this one-sided difference is chosen because it leads to a reduction in the slope and suppresses oscillations. However, the idea behind this method fails at extrema because the derivative vanishes. Furthermore, the one-sided differences have opposite signs and this non-constant multiple bounds the centered difference. In the original van Leer and PPM limiters, the solution is to simply drop the order of the method to $1^{st}$ order. 
\\

\noindent
In the approach used in this report, one changes the limiters at extrema, and only at extrema, by using comparisons of different estimates of the $2^{nd}$ derivatives as a basis for whether to limit the underlying linear scheme. If the solution is smooth at the extremum, then all of the estimates of the $2^{nd}$ derivative are comparable and the limiter leaves the underlying linear scheme unchanged. Discontinuities, underresolved gradients, and high-wavenumber oscillations are detected either by one of the estimates of the $2^{nd}$ derivative being much smaller than the others or by the various estimates of the $2^{nd}$ derivatives changing sign. Either effect triggers a nontrivial limiting of the interpolating function in the cell and a resulting suppression of oscillations.

\subsection{Extremum-Preserving van Leer Limiter}
\noindent
The extremum-preserving van Leer limiter parallels the conventional van Leer limiter:

\begin{enumerate}
\item Compute one-sided and centered differences as well as one-sided differences that are an additional spatial step-size away:
\begin{eqnarray}
\Delta_{--} a_i &=& a_{i-1} - a_{i-2} , \\
\Delta_{-} a_i  &=& a_{i} - a_{i-1} , \\
\Delta_{c} a_i  &=& \frac{1}{2} (a_{i+1} - a_{i-1}) , \\
\Delta_{+} a_i  &=& a_{i+1} - a_{i} , \\
\Delta_{++} a_i &=& a_{i+2} - a_{i+1} .
\end{eqnarray}

\item Test for extrema. An extremum is defined if the following condition is satisfied:
\begin{equation}
\min(\Delta_{-} ~ a_i \Delta_{+} a_i, ~ \Delta_{--} a_i ~ \Delta_{++} a_i) < 0 .
\end{equation}

\item If the above extremum condition is not satisfied, then limiting follows the conventional van Leer method:
\begin{eqnarray}
\Delta_{\lim} a_i &=& 2 \min(|\Delta_{-} a_i|,|\Delta_{+} a_i|) , \\
\mathcal{S}       &=& \textrm{sign}(\Delta_{c} a_i) , \\
\Delta a_i        &=& \min(|\Delta_{c} a_i|,|\Delta_{\lim} a_i|) \mathcal{S} .
\end{eqnarray}

\item If the above extremum condition is satisfied, then:

\begin{itemize}
\item Compute one-sided and centered $2^{nd}$ derivatives:
\begin{eqnarray}
\mathcal{D}^{2}_{-} a_{i} &=& \frac{1}{h^2} \left( a(ih)     - 2 a((i-1)h) + a((i-2)h) \right) , \\
\mathcal{D}^{2}_{c} a_{i} &=& \frac{1}{h^2} \left( a((i+1)h) - 2 a(ih)     + a((i-1)h) \right) , \\
\mathcal{D}^{2}_{+} a_{i} &=& \frac{1}{h^2} \left( a((i+2)h) - 2 a((i+1)h) + a(ih)     \right) .
\end{eqnarray}

\item Find the minimum $2^{nd}$ derivative over the five-cells in question:
\begin{eqnarray} 
\mathcal{S}^{2}              &=& \textrm{sign}(\mathcal{D}^{2}_{c} a_{i}) , \\
\mathcal{D}^{2}_{\lim} a_{i} &=& \min(|\mathcal{D}^{2}_{c} a_{i}|,\max(\mathcal{S}^{2} \mathcal{D}^{2}_{-} a_{i},0),\max(\mathcal{S}^{2} \mathcal{D}^{2}_{+} a_{i},0)) .
\end{eqnarray}

\item Apply the modified van Leer limiter at the extremum:
\begin{eqnarray}
\Delta_{\lim} a_i &=& \left \{  \begin{array}{ll} 
     \min \left( C_{VL} \frac{3 h^{2}}{2} \mathcal{D}^{2}_{\lim} a_{i}, 2 |\Delta_{-} a_i| \right) & \mathcal{S}^{2} \Delta_{c} a_{i} < 0 \\
     \min \left( C_{VL} \frac{3 h^{2}}{2} \mathcal{D}^{2}_{\lim} a_{i}, 2 |\Delta_{+} a_i| \right) & \textrm{else}
     \end{array} \right . \\
\mathcal{S}       &=& \textrm{sign}(\Delta_{c} a_i) , \\
\Delta a_i        &=& \min(|\Delta_{c} a_i|,|\Delta_{\lim} a_i|) \mathcal{S} .
\end{eqnarray}

\end{itemize}

\end{enumerate}

\noindent
$\mathcal{D}$ designates a derivative while $\Delta$ designates a difference. Derivatives and differences are similar operators that can be transformed back-and-forth when one considers the relevant Taylor expansion and multiplies/divides each operator by factors of $h$. $C_{VL}$ is a constant that is independent of the mesh spacing and as $C_{VL} \rightarrow 0$, the extremum-preserving van Leer limiter reduces to the conventional van Leer limiter. For most calculations, $C_{VL} = 1.25$. 
\\

\noindent
How one arrives at the tighter bound of $\frac{3 h^2}{2} \mathcal{D}^{2}_{\lim} a_{i}$ for peak height in the extremum-preserving van Leer limiter can be understood by considering the following example. Assume that one is at a local maximum such that $\frac{d^2 a}{dx^2}|_{ih} < 0$ and $\frac{da}{dx}|_{ih} < 0$. The van Leer limiting condition bounds the derivative on the high (right) edge of the cell:
\begin{equation}
\Delta a_{i} \geq 2 (a_{i+1} -a_{i}) .
\end{equation}

\noindent
When looking for a bound on $\Delta a_{i}$ from the low (left) edge of the cell, an extremum near $i$ implies that:
\begin{equation}
0 \leq a((i-1)h) - a((i-2)h) = h \frac{da}{dx}|_{(i-\frac{3}{2})h} + O(h^3) .
\end{equation}

\noindent
Therefore, one arrives at the following bound on $\Delta a_{i}$:
\begin{equation}
\Delta a_{i} = h \frac{da}{dx}|_{ih} + O(h^3) = h \frac{da}{dx}|_{(i-\frac{3}{2})h} + \frac{3 h^2}{2} \frac{d^2 a}{dx^2} + O(h^3) \geq \frac{3 h^2}{2} \frac{d^2 a}{dx^2} + O(h^3) .
\end{equation}

\noindent
These bounds are summarized in the following inequality:
\begin{equation}
0 \geq \Delta a_{i} \geq C_{VL} \frac{3 h^2}{2} \frac{d^2 a}{dx^2}, ~~~ C_{VL} > 1 ,
\end{equation}

\noindent
where the $3/2$ comes from the condition that $(i-\frac{3}{2})h$ is the nearest face-centered point to the cell-centered point $ih$ for which one can unequivocally assert that $h \frac{da}{dx}|_{(i-\frac{3}{2})h} \geq O(h)$.

\subsection{Extremum-Preserving Parabolic Profile Limiter}
\noindent
If the piecewise parabolic reconstruction is done without using van Leer limited differences, then one has to include an additional step that limits extrema at cell faces such that $a_{i+\frac{1}{2}}$ lies between adjacent cell averages. Van Leer limiting automatically enforces this constraint.
\begin{itemize}
\item Test for extremum at cell faces: 
\begin{equation}
(a_{i+\frac{1}{2}}-a_{i})(a_{i+1}-a_{i+\frac{1}{2}}) < 0 .
\end{equation}

\noindent
If no extremum is found, then proceed without any adjustment to $a_{i+\frac{1}{2}}$.

\item If an extremum is found, compute one-sided and centered $2^{nd}$ derivatives:
\begin{eqnarray}
\mathcal{D}^{2}_{-} a_{i+\frac{1}{2}} &=& \frac{1}{h^2} \left( a((i+1)h) - 2 a(ih) + a((i-1)h) \right)           , \\ 
\mathcal{D}^{2}_{c} a_{i+\frac{1}{2}} &=& \frac{3}{h^2} \left( a((i+1)h) - 2 a((i+\frac{1}{2})h) + a(ih) \right) , \\
\mathcal{D}^{2}_{+} a_{i+\frac{1}{2}} &=& \frac{1}{h^2} \left( a((i+2)h) - 2 a((i+1)h) + a(ih) \right)           .
\end{eqnarray}

\item Find the minimum difference with respect to $a_{i+\frac{1}{2}}$ that is greater than zero:
\begin{eqnarray}
\mathcal{S}^{2}_{i+\frac{1}{2}}          &=& \textrm{sign}(\mathcal{D}^{2}_{c} a_{i+\frac{1}{2}}) , \\
\mathcal{D}^{2}_{\lim} a_{i+\frac{1}{2}} &=& \max ( \min ( C_{PPM} \mathcal{S}^{2}_{i+\frac{1}{2}} \mathcal{D}^{2}_{-} a_{i+\frac{1}{2}} , \mathcal{S}^{2}_{i+\frac{1}{2}} \mathcal{D}^{2}_{c} a_{i+\frac{1}{2}} , \nonumber \\
                                         & & ~~~~~~~~~~~~~ C_{PPM} \mathcal{S}^{2}_{i+\frac{1}{2}} \mathcal{D}^{2}_{+} a_{i+\frac{1}{2}} ) , 0 ) .
\end{eqnarray}

\item Adjust $a_{i+\frac{1}{2}}$ according to:
\begin{equation}
a_{i+\frac{1}{2}} = \frac{1}{2}(a_{i+1}+a_{i}) - \frac{1}{6} \mathcal{D}^{2}_{\lim} a_{i+\frac{1}{2}} .
\end{equation}

\end{itemize}

\noindent 
The above formulas are arrived at by considering the centered $2^{nd}$ derivative:
\begin{equation}
\mathcal{D}^{2}_{c} a_{i+\frac{1}{2}} = \frac{1}{(h/2)^2} \left( a((i+1)h) - 2 a((i+\frac{1}{2})h) + a(ih) \right) .
\end{equation}

\noindent
The spatial resolution of $h/2$ enters because one is considering differences between $a(ih),~a((i+\frac{1}{2})h),~a((i+1)h)$. By substituting in the following expressions:
\begin{eqnarray}
a_{i}   &=& a(ih)     - \frac{h^2}{24} \mathcal{D}^{2}_{c} a_{i}   + O(h^4) , \\
a_{i+1} &=& a((i+1)h) - \frac{h^2}{24} \mathcal{D}^{2}_{c} a_{i+1} + O(h^4) ,
\end{eqnarray}

\noindent
which convert point-valued to average-valued quantities, one arrives at:
\begin{equation}
\frac{h^2}{4} \mathcal{D}^{2}_{c} a_{i+\frac{1}{2}} = a_{i+1} - 2 a((i+\frac{1}{2})h) + a_{i} - \frac{h^2}{12} \mathcal{D}^{2}_{c} a_{i+\frac{1}{2}} + O(h^4) .
\end{equation}

\noindent
Rearranging the above equation gives the expression:
\begin{equation}
\mathcal{D}^{2}_{c} a_{i+\frac{1}{2}} = \frac{3}{h^2} \left( a_{i+1} - 2 a((i+\frac{1}{2})h) + a_{i} \right) .
\end{equation}

\noindent
After finding the minimum $2^{nd}$ derivative $\mathcal{D}^{2}_{\lim} a_{i+\frac{1}{2}}$, one can again use the above equation to assign a value to $a_{i+\frac{1}{2}}$:
\begin{equation}
a_{i+\frac{1}{2}} = \frac{1}{2} \left( a_{i+1} + a_{i} \right) - \frac{1}{6} \mathcal{D}^{2}_{\lim} a_{i+\frac{1}{2}} .
\end{equation}

\noindent
Lastly, $a_{\pm}=a_{i\pm\frac{1}{2}}$ and one proceeds with limiting $\alpha_{\pm}$.
\\

\noindent
Now given the higher-order reconstruction of $a_{\pm}$ such that $\alpha_{\pm} = a_{\pm} - a_{i}$, adjust $\alpha_{\pm}$ according to the following cases:

\begin{enumerate}
\item $\alpha_{+} \alpha_{-} \geq 0 ~~ \big\| ~~ (a_{i+1} - a_{i})(a_{i} - a_{i-1}) \leq 0$ 

\begin{itemize}
\item Compute the $2^{nd}$ derivative with respect to $\sigma$ from the the interpolation polynomial described in the original Piecewise Parabolic Method:
\begin{equation}
\mathcal{D}^{2}_{PPM} a_{i} = \frac{d^2 a_i}{d^2 \sigma} = -2 a_{6} = 6 (\alpha_{+} + \alpha_{-}) .
\end{equation}

\noindent
It is important to note that this $2^{nd}$ derivative with respect to $\sigma$ is also a $2^{nd}$ order difference for $a_{i}$. Furthermore, this difference represents the maximum difference that can occur across a cell given a parabolic profile.

\item Compute one-sided and centered $2^{nd}$ order derivatives:
\begin{eqnarray}
\mathcal{D}^{2}_{-} a_{i} &=& \frac{1}{h^2} \left( a(ih)     - 2 a((i-1)h) + a((i-2)h) \right) , \\
\mathcal{D}^{2}_{c} a_{i} &=& \frac{1}{h^2} \left( a((i+1)h) - 2 a(ih)    + a((i-1)h)  \right) , \\
\mathcal{D}^{2}_{+} a_{i} &=& \frac{1}{h^2} \left( a((i+2)h) - 2 a((i+1)h) + a(ih)     \right) .
\end{eqnarray}

\item Find the minimum difference that is greater than zero, given the $2^{nd}$ derivatives over the five-cells in question as well as the above difference that was derived from the interpolation polynomial:
\begin{eqnarray}
\mathcal{S}^{2}_{PPM}        &=& \textrm{sign}(\mathcal{D}^{2}_{PPM} a_i) , \\
\mathcal{D}^{2}_{\lim} a_{i} &=& \max ( \min ( \mathcal{S}^{2}_{PPM} \mathcal{D}^{2}_{PPM} a_i , C_{PPM} \mathcal{S}^{2}_{PPM} \mathcal{D}^{2}_{-} a_{i} , \nonumber \\
                             & & ~~~~~~~~~~~~~ C_{PPM} \mathcal{S}^{2}_{PPM} \mathcal{D}^{2}_{c} a_i , C_{PPM} \mathcal{S}^{2}_{PPM} \mathcal{D}^{2}_{+} a_i ) , 0 ) .
\end{eqnarray}

\item Adjust $\alpha_{\pm}$ according to $\mathcal{D}^{2}_{\lim} a_{i}$ and $\mathcal{D}^{2}_{PPM} a_{i}$:
\begin{equation}
\alpha_{\pm} \rightarrow \alpha_{\pm} \frac{ | \mathcal{D}^{2}_{\lim} a_{i} | }{ | \mathcal{D}^{2}_{PPM} a_{i} | } .
\end{equation}

\end{itemize}

\item $\alpha^{2}_{+} > 4 \alpha^{2}_{-}$

\begin{itemize}
\item Compute the maximum value of $\alpha_{-}$ over a given cell:
\begin{equation}
\alpha^{\max}_{-} = \frac{-\alpha^{2}_{+}}{4(\alpha_{+}+\alpha_{-})} .
\end{equation}

\noindent 
This formula is arrived at by averaging the interpolation polynomial from the original Piecewise Parabolic Method on the left side of cell $i$:
\begin{eqnarray}
f_{-} &=& \frac{1}{\sigma} \int^{\sigma}_{0} a(\xi) \xi \\
      &=& a_{-} + \frac{\sigma}{2} \left( \delta_{\pm} a + \left( 1 - \frac{2 \sigma}{3} a_{6} \right) \right) \\
      &=& a_{-} + \frac{\sigma}{2} \left( \alpha_{+} - \alpha_{-} \right) - \frac{3 \sigma}{2} \left( 1 - \frac{2 \sigma}{3} \right) (\alpha_{+} + \alpha_{-}) .
\end{eqnarray}

\noindent
Maximize $f_{-}$ with respect to $\sigma$:
\begin{equation}
\frac{d f_{-}}{d \sigma} = 0 \Rightarrow \sigma_{\max} = \frac{\alpha_{+} + 2 \alpha_{-}}{2 (\alpha_{+} + \alpha_{-})} .
\end{equation}

\noindent
Therefore:
\begin{equation}
\alpha^{\max}_{-} = f_{-} (\sigma_{\max}) - a_{i} = \frac{-\alpha^{2}_{+}}{4(\alpha_{+}+\alpha_{-})} .
\end{equation}

\item Preserve monotonicity by ensuring that the following condition is just satisfied:
\begin{equation}
I_{-} \leq a_{i-1} - a_{i} \Rightarrow I_{-} = a_{i-1} - a_{i} = \alpha^{\max}_{-} .
\end{equation}

\item Solve the above equation for $\alpha_{+}$:
\begin{eqnarray}
\mathcal{S}^{-} &=& \textrm{sign}(\alpha_{-}) , \\
\alpha_{+}      &=& -2 I_{-} - 2 \mathcal{S}^{-} \left( I^{2}_{-} - \alpha_{-} I_{-} \right)^{1/2} .
\end{eqnarray}

\end{itemize}

\item $\alpha^{2}_{-} > 4 \alpha^{2}_{+}$

\begin{itemize}
\item Compute the maximum value of $\alpha_{+}$ over a given cell:
\begin{equation}
\alpha^{\max}_{+} = \frac{-\alpha^{2}_{-}}{4(\alpha_{+}+\alpha_{-})} .
\end{equation}

\noindent 
This formula is arrived at by averaging the interpolation polynomial from the original Piecewise Parabolic Method on the right side of cell $i$:
\begin{eqnarray}
f_{+} &=& \frac{1}{\sigma} \int^{1}_{1-\sigma} a(\xi) \xi \\
      &=& a_{+} - \frac{\sigma}{2} \left( \delta_{\pm} a - \left( 1 - \frac{2 \sigma}{3} a_{6} \right) \right) \\
      &=& a_{+} - \frac{\sigma}{2} \left( \alpha_{+} - \alpha_{-} \right) - \frac{3 \sigma}{2} \left( 1 - \frac{2 \sigma}{3} \right) (\alpha_{+} + \alpha_{-}) .
\end{eqnarray}

\noindent
Maximize $f_{+}$ with respect to $\sigma$:
\begin{equation}
\frac{d f_{+}}{d \sigma} = 0 \Rightarrow \sigma_{\max} = \frac{2 \alpha_{+} + \alpha_{-}}{2 (\alpha_{+} + \alpha_{-})} .
\end{equation}

\noindent
Therefore:
\begin{equation}
\alpha^{\max}_{+} = f_{+} (\sigma_{\max}) - a_{i} = \frac{-\alpha^{2}_{-}}{4(\alpha_{+}+\alpha_{-})} .
\end{equation}

\item Preserve monotonicity by ensuring that the following condition is just satisfied:
\begin{equation}
I_{+} \geq a_{i+1} - a_{i} \Rightarrow I_{+} = a_{i+1} - a_{i} = \alpha^{\max}_{+} .
\end{equation}

\item Solve the above equation for $\alpha_{-}$:
\begin{eqnarray}
\mathcal{S}^{+} &=& \textrm{sign}(\alpha_{+}) , \\
\alpha_{-}      &=& -2 I_{+} - 2 \mathcal{S}^{+} \left( I^{2}_{+} - \alpha_{+} I_{+} \right)^{1/2} .
\end{eqnarray}

\end{itemize}

\item Reconstruct $a_{\pm}$ given the adjusted values for $\alpha_{\pm}$.
\end{enumerate}


\section{Numerical Results}
\noindent
Results are presented for 1D scalar advection. The standard test problems were employed to demonstrate improvements in accuracy \cite{zalesak:1984}. The following parameters were used in all test problems:
\begin{equation}
\sigma = 0.2, ~~ x \in [0,1], ~~  t = 10, \lambda = 1 .
\end{equation}

\noindent
Periodic boundary conditions were used with the following number of ghost cells:
\begin{eqnarray}
\textrm{$4^{th}$ Order Reconstruction} & \Rightarrow & 3 \textrm{ ghost cells per side} \\
\textrm{$6^{th}$ Order Reconstruction} & \Rightarrow & 4 \textrm{ ghost cells per side} . 
\end{eqnarray}

\noindent
The test problems were defined by the following initial conditions:
\begin{eqnarray}
\textrm{Gaussian Wave:    } a & = & \exp \left( -256 (x-0.5)^{2} \right) \\
\textrm{Semi-Circle Wave: } a & = & \left\{ \begin{array}{ll}
                                         \left( 0.25^{2} - (x-0.5)^{2} \right) & 0.25 < x < 0.75 \\
                                         0 & \rm{else} \end{array} \right . \\
\textrm{Square Wave:      } a & = & \left\{ \begin{array}{ll}
                                         1 & 0.25 < x < 0.75 \\
                                         0 & \rm{else} \end{array} \right .
\end{eqnarray}

\noindent
The conventional initialization for $2^{nd}$ order accurate methods is obtained by approximating the average over a cell by the value at the center of the cell, i.e., the midpoint rule for integrals. However, that initialization is only $2^{nd}$ order accurate, and less accurate than what one expects for PPM when applied to the advection equation. For that reason, one uses a $4^{th}$ order accurate approximation to the cell average, following  \cite{baradColella:2006}:
\begin{equation}
a_{i} = a(i h) + \frac{h^2}{24} \mathcal{D}^{2}_{c} a_i , ~~ \mathcal{D}^{2}_{c} a_i = \frac{1}{h^2} \left( a(i+1)h) - 2 a(i h) + a((i-1)h) \right) .
\end{equation}

\noindent
This $4^{th}$ order correction is only administered when defining the initial conditions. The corresponding boundary conditions are applied and one begins advancing the temporal loop. For smooth solutions away from extrema, PPM is $3^{rd}$ order accurate for linear advection and $4^{th}$ order accurate in the limit of vanishing CFL number $\sigma \rightarrow 0$. The following definitions for the n-norms and convergence rates are used throughout this note. Given the numerical solution $a^r$ at resolution $r$ and the analytic solution $u$, the error at a given point $i$ is:
\begin{equation}
\epsilon^{r}_{i} = a^{r}_{i} - u_{i}.
\end{equation}

\noindent
The 1-norm and $\infty$-norm of the error are:
\begin{equation}
L_{1} = \parallel \epsilon \parallel_{1} = \sum_{i} | \epsilon_{i} | \Delta x , ~~~~ L_{\infty} = \parallel \epsilon \parallel_{\infty} = \max_{i} | \epsilon_{i} | .
\end{equation}

\noindent
The convergence rate is measured using Richardson extrapolation:
\begin{equation}
R_n = \frac{ \textrm{ln} \left( L_n(\epsilon^r) / L_n(\epsilon^{r+1}) \right) }{ \textrm{ln} \left( \Delta x^{r} / \Delta x^{r+1} \right) } .
\end{equation}

\clearpage

\subsection{$4^{th}$ Order Reconstruction}
~

\begin{center}
\begin{tabular*}{1.0\textwidth}{@{\extracolsep{\fill}}lrllll}
\hline
\multicolumn{6}{l}{$4^{th}$ Order, Gaussian Wave} \\
\hline
  & $N_{cell}$ & $L_1$ & $R_1$ & $L_{\infty}$ & $R_{\infty}$ \\
\hline
\textrm{No Limiter} & & & & & \\
  &  32 & 8.0E-2 & -   & 3.5E-1 & -   \\
  &  64 & 2.6E-2 & 1.6 & 1.5E-1 & 1.2 \\
  & 128 & 3.2E-3 & 3.0 & 2.8E-2 & 2.5 \\
  & 256 & 3.1E-4 & 3.4 & 3.0E-3 & 3.2 \\
\hline
\textrm{PPM Limiting $~C_{PPM}=0$} & & & & & \\
  &  32 & 7.5E-2 & -   & 4.7E-1 & -   \\
  &  64 & 2.7E-2 & 1.5 & 2.6E-1 & 0.9 \\
  & 128 & 7.8E-3 & 1.8 & 9.9E-2 & 1.4 \\
  & 256 & 1.3E-3 & 2.6 & 3.1E-2 & 1.7 \\
\hline
\textrm{PPM Limiting $~C_{PPM}=1.25$} & & & & & \\
  &  32 & 5.5E-2 & -   & 3.7E-1 & -   \\
  &  64 & 1.6E-2 & 1.8 & 1.4E-1 & 1.4 \\
  & 128 & 3.2E-3 & 2.3 & 2.8E-2 & 2.4 \\
  & 256 & 3.1E-4 & 3.4 & 3.0E-3 & 3.2 \\
\hline
\textrm{VL+PPM Limiting $~C_{VL}, C_{PPM}=1.25$} & & & & & \\
  &  32 & 5.7E-2 & -   & 3.8E-1 & -   \\
  &  64 & 1.6E-2 & 1.8 & 1.5E-1 & 1.4 \\
  & 128 & 3.3E-3 & 2.3 & 2.8E-2 & 2.4 \\
  & 256 & 3.1E-3 & 3.4 & 3.0E-3 & 3.2 \\
\hline
\end{tabular*}
\end{center}
\pagebreak

\begin{center}
\begin{tabular*}{1.0\textwidth}{@{\extracolsep{\fill}}lrllll}
\hline
\multicolumn{6}{l}{$4^{th}$ Order, Semi-Circle Wave} \\
\hline
  & $N_{cell}$ & $L_1$ & $R_1$ & $L_{\infty}$ & $R_{\infty}$ \\
\hline
\textrm{No Limiter} & & & & & \\
  &  32 & 1.2E-2 & -   & 5.2E-2 & -   \\
  &  64 & 5.9E-3 & 1.0 & 4.1E-2 & 0.3 \\
  & 128 & 2.6E-3 & 1.2 & 3.2E-2 & 0.3 \\
  & 256 & 1.1E-3 & 1.2 & 2.5E-2 & 0.3 \\
\hline
\textrm{PPM Limiting $~C_{PPM}=0$} & & & & & \\
  &  32 & 8.1E-3 & -   & 4.1E-2 & -   \\
  &  64 & 4.7E-3 & 0.8 & 3.4E-2 & 0.3 \\
  & 128 & 2.1E-3 & 1.2 & 2.7E-2 & 0.3 \\
  & 256 & 9.0E-4 & 1.2 & 2.1E-2 & 0.3 \\
\hline
\textrm{PPM Limiting $~C_{PPM}=1.25$} & & & & & \\
  &  32 & 8.7E-3 & -   & 4.2E-2 & -   \\
  &  64 & 4.5E-3 & 0.9 & 3.4E-2 & 0.3 \\
  & 128 & 2.0E-3 & 1.2 & 2.7E-2 & 0.3 \\
  & 256 & 8.9E-4 & 1.2 & 2.1E-2 & 0.3 \\
\hline
\textrm{VL+PPM Limiting $~C_{VL}, C_{PPM}=1.25$} & & & & & \\
  &  32 & 7.7E-3 & -   & 4.1E-2 & -   \\
  &  64 & 4.1E-3 & 0.9 & 3.3E-2 & 0.3 \\
  & 128 & 1.8E-3 & 1.2 & 2.7E-2 & 0.3 \\
  & 256 & 8.0E-4 & 1.2 & 2.1E-2 & 0.3 \\
\hline
\end{tabular*}
\end{center}
\pagebreak

\begin{center}
\begin{tabular*}{1.0\textwidth}{@{\extracolsep{\fill}}lrllll}
\hline
\multicolumn{6}{l}{$4^{th}$ Order, Square Wave} \\
\hline
  & $N_{cell}$ & $L_1$ & $R_1$ & $L_{\infty}$ & $R_{\infty}$ \\
\hline
\textrm{No Limiter} & & & & & \\
  &  32 & 1.1E-1 & -   & 4.5E-1 & -    \\
  &  64 & 7.6E-2 & 0.5 & 4.7E-1 & -0.1 \\
  & 128 & 4.5E-2 & 0.8 & 4.8E-1 &  0.0 \\
  & 256 & 2.5E-2 & 0.8 & 4.9E-1 &  0.0 \\
\hline
\textrm{PPM Limiting $~C_{PPM}=0$} & & & & & \\
  &  32 & 9.1E-2 & -   & 4.2E-1 & -    \\
  &  64 & 5.2E-2 & 0.8 & 4.3E-1 &  0.0 \\
  & 128 & 3.0E-2 & 0.8 & 4.4E-1 &  0.0 \\
  & 256 & 1.7E-2 & 0.8 & 4.5E-1 &  0.0 \\
\hline
\textrm{PPM Limiting $~C_{PPM}=1.25$} & & & & & \\
  &  32 & 9.0E-2 & -   & 4.2E-1 & -    \\
  &  64 & 5.2E-2 & 0.8 & 4.3E-1 &  0.0 \\
  & 128 & 3.0E-2 & 0.8 & 4.4E-1 &  0.0 \\
  & 256 & 1.7E-2 & 0.8 & 4.5E-1 &  0.0 \\
\hline
\textrm{VL+PPM Limiting $~C_{VL}, C_{PPM}=1.25$} & & & & & \\
  &  32 & 8.0E-2 & -   & 4.0E-1 & -    \\
  &  64 & 4.6E-2 & 0.8 & 4.1E-1 &  0.0 \\
  & 128 & 2.7E-2 & 0.8 & 4.2E-1 &  0.0 \\
  & 256 & 1.6E-2 & 0.8 & 4.3E-1 &  0.0 \\
\hline
\end{tabular*}
\end{center}
\pagebreak

\subsection{$6^{th}$ Order Reconstruction}
~

\begin{center}
\begin{tabular*}{1.0\textwidth}{@{\extracolsep{\fill}}lrllll}
\hline
\multicolumn{6}{l}{$6^{th}$ Order, Gaussian Wave} \\
\hline
  & $N_{cell}$ & $L_1$ & $R_1$ & $L_{\infty}$ & $R_{\infty}$ \\
\hline
\textrm{No Limiter} & & & & & \\
  &  32 & 5.0E-2 & -   & 2.6E-1 & -   \\
  &  64 & 1.3E-2 & 2.0 & 9.8E-2 & 1.4 \\
  & 128 & 2.0E-3 & 2.7 & 1.8E-2 & 2.4 \\
  & 256 & 2.6E-4 & 2.9 & 2.5E-3 & 2.9 \\
\hline
\textrm{PPM Limiting $~C_{PPM}=0$} & & & & & \\
  &  32 & 6.7E-2 & -   & 4.4E-1 & -   \\
  &  64 & 2.3E-2 & 1.5 & 2.3E-1 & 0.9 \\
  & 128 & 5.3E-3 & 2.1 & 8.7E-2 & 1.4 \\
  & 256 & 9.2E-4 & 2.5 & 2.7E-2 & 1.7 \\
\hline
\textrm{PPM Limiting $~C_{PPM}=1.25$} & & & & & \\
  &  32 & 4.1E-2 & -   & 2.9E-1 & -   \\
  &  64 & 1.1E-2 & 1.9 & 9.7E-2 & 1.6 \\
  & 128 & 2.0E-3 & 2.5 & 1.8E-2 & 2.4 \\
  & 256 & 2.6E-4 & 2.9 & 2.5E-3 & 2.9 \\
\hline
\textrm{VL+PPM Limiting $~C_{VL}, C_{PPM}=1.25$} & & & & & \\
  &  32 & 4.4E-2 & -   & 3.2E-1 & -   \\
  &  64 & 1.2E-2 & 1.9 & 1.1E-1 & 1.5 \\
  & 128 & 2.0E-3 & 2.6 & 1.8E-2 & 2.6 \\
  & 256 & 2.6E-3 & 2.9 & 2.6E-3 & 2.9 \\
\hline
\end{tabular*}
\end{center}
\pagebreak

\begin{center}
\begin{tabular*}{1.0\textwidth}{@{\extracolsep{\fill}}lrllll}
\hline
\multicolumn{6}{l}{$6^{th}$ Order, Semi-Circle Wave} \\
\hline
  & $N_{cell}$ & $L_1$ & $R_1$ & $L_{\infty}$ & $R_{\infty}$ \\
\hline
\textrm{No Limiter} & & & & & \\
  &  32 & 8.4E-3 & -   & 4.2E-2 & -   \\
  &  64 & 3.7E-3 & 1.2 & 3.4E-2 & 0.3 \\
  & 128 & 1.6E-3 & 1.2 & 2.7E-2 & 0.3 \\
  & 256 & 7.4E-4 & 1.1 & 2.2E-2 & 0.3 \\
\hline
\textrm{PPM Limiting $~C_{PPM}=0$} & & & & & \\
  &  32 & 7.5E-3 & -   & 3.6E-2 & -   \\
  &  64 & 3.4E-3 & 1.2 & 3.0E-2 & 0.3 \\
  & 128 & 1.4E-3 & 1.2 & 2.4E-2 & 0.3 \\
  & 256 & 6.3E-4 & 1.2 & 1.9E-2 & 0.3 \\
\hline
\textrm{PPM Limiting $~C_{PPM}=1.25$} & & & & & \\
  &  32 & 7.3E-3 & -   & 3.7E-2 & -   \\
  &  64 & 3.2E-3 & 1.2 & 3.0E-2 & 0.3 \\
  & 128 & 1.4E-3 & 1.2 & 2.4E-2 & 0.3 \\
  & 256 & 6.1E-4 & 1.2 & 1.9E-2 & 0.3 \\
\hline
\textrm{VL+PPM Limiting $~C_{VL}, C_{PPM}=1.25$} & & & & & \\
  &  32 & 7.1E-3 & -   & 3.6E-2 & -   \\
  &  64 & 2.9E-3 & 1.3 & 2.9E-2 & 0.3 \\
  & 128 & 1.2E-3 & 1.3 & 2.3E-2 & 0.3 \\
  & 256 & 5.0E-4 & 1.2 & 1.9E-2 & 0.3 \\
\hline
\end{tabular*}
\end{center}
\pagebreak

\begin{center}
\begin{tabular*}{1.0\textwidth}{@{\extracolsep{\fill}}lrllll}
\hline
\multicolumn{6}{l}{$6^{th}$ Order, Square Wave} \\
\hline
  & $N_{cell}$ & $L_1$ & $R_1$ & $L_{\infty}$ & $R_{\infty}$ \\
\hline
\textrm{No Limiter} & & & & & \\
  &  32 & 9.8E-2 & -   & 4.1E-1 & -    \\
  &  64 & 5.6E-2 & 0.8 & 4.2E-1 & 0.0 \\
  & 128 & 3.2E-2 & 0.8 & 4.3E-1 & 0.0 \\
  & 256 & 1.9E-2 & 0.8 & 4.4E-1 & 0.0 \\
\hline
\textrm{PPM Limiting $~C_{PPM}=0$} & & & & & \\
  &  32 & 7.8E-2 & -   & 3.9E-1 & -    \\
  &  64 & 4.5E-2 & 0.8 & 4.1E-1 & 0.0 \\
  & 128 & 2.6E-2 & 0.8 & 4.2E-1 & 0.0 \\
  & 256 & 1.5E-2 & 0.8 & 4.3E-1 & 0.0 \\
\hline
\textrm{PPM Limiting $~C_{PPM}=1.25$} & & & & & \\
  &  32 & 7.7E-2 & -   & 3.9E-1 & -    \\
  &  64 & 4.4E-2 & 0.8 & 4.0E-1 & -0.1 \\
  & 128 & 2.6E-2 & 0.8 & 4.2E-1 &  0.0 \\
  & 256 & 1.5E-2 & 0.8 & 4.3E-1 &  0.0 \\
\hline
\textrm{VL+PPM Limiting $~C_{VL}, C_{PPM}=1.25$} & & & & & \\
  &  32 & 6.5E-2 & -   & 3.6E-1 & -    \\
  &  64 & 3.7E-2 & 0.8 & 3.8E-1 & -0.1 \\
  & 128 & 2.1E-2 & 0.8 & 4.0E-1 & -0.1 \\
  & 256 & 1.2E-2 & 0.8 & 4.2E-1 & -0.1 \\
\hline
\end{tabular*}
\end{center}
\pagebreak

\subsection{Comparing Conventional and Extremum-Preserving Limiters}
\begin{figure}
\begin{center}
\includegraphics[width=4in,angle=0]{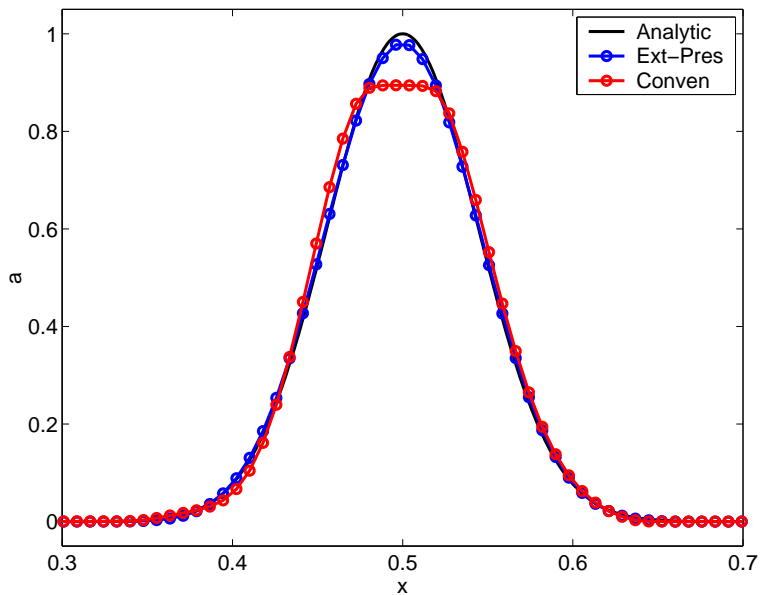}
\caption{\label{fig:gaussian}Extremum-Preserving vs Conventional Limiter. Gaussian Pulse, $N_{cell}=128$, $t=10$ periods, $CFL=0.2$, $\lambda=1$.}
\end{center}
\end{figure}

\begin{figure}
\begin{center}
\includegraphics[width=4in,angle=0]{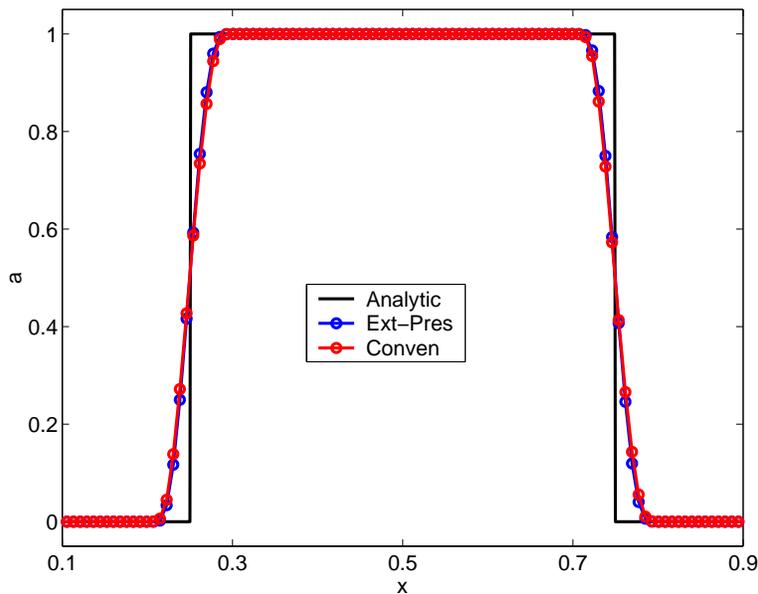}
\caption{\label{fig:square}Extremum-Preserving vs Conventional Limiter. Square Wave, $N_{cell}=128$, $t=10$ periods, $CFL=0.2$, $\lambda=1$.}
\end{center}
\end{figure}

\noindent
This section compares the conventional limiter and extremum-preserving limiter when used in PPM for the one-dimensional scalar advection tests. The conventional limiter uses $2^{nd}$ order van Leer limited differences to compute spatially $4^{th}$ order accurate extrapolated face-centered values \cite{colellaWoodward:1984,millerColella:2002}. The extremum-preserving limiter employs $6^{th}$ order accurate piecewise parabolic reconstruction without using the van Leer limited differences \cite{colellaSekora:2007}. As expected, the extremum-preserving limiters significantly reduce the error for the Gaussian wave (G), reduce the error only slightly for the semi-circle wave (SC), and does not reduce the error for the square wave (S). This comparison is shown in Figures \ref{fig:gaussian} and \ref{fig:square}.
\\

\begin{center}
\begin{tabular*}{1.0\textwidth}{@{\extracolsep{\fill}}rcccccccc}
\hline
\multicolumn{9}{l}{Conventional Limiter} \\
\hline
$N_{cell}$ & G $L_1$ & $R_1$ & G $L_{\infty}$ & $R_{\infty}$ & SC $L_1$ & $R_1$ & S $L_1$ & $R_1$ \\
\hline
 32 & 7.6E-2 & -   & 4.8E-1 & -   & 7.8E-3 & -   & 8.4E-2 & -   \\
 64 & 2.7E-2 & 1.5 & 2.7E-1 & 0.8 & 4.3E-3 & 0.9 & 4.8E-2 & 0.8 \\
128 & 7.7E-3 & 1.8 & 1.0E-1 & 1.4 & 1.9E-3 & 1.2 & 2.8E-2 & 0.8 \\
256 & 1.3E-3 & 2.6 & 3.1E-2 & 1.7 & 8.3E-4 & 1.2 & 1.6E-2 & 0.8 \\
\hline
\end{tabular*}
\end{center}
~
\\

\begin{center}
\begin{tabular*}{1.0\textwidth}{@{\extracolsep{\fill}}rcccccccc}
\hline
\multicolumn{9}{l}{Extremum-Preserving Limiter ($C=1.25$)} \\
\hline
$N_{cell}$ & G $L_1$ & $R_1$ & G $L_{\infty}$ & $R_{\infty}$ & SC $L_1$ & $R_1$ & S $L_1$ & $R_1$ \\
\hline
 32 & 4.1E-2 & -   & 2.9E-1 & -   & 7.3E-3 & -   & 7.7E-2 & -   \\
 64 & 1.1E-2 & 1.9 & 9.7E-2 & 1.6 & 3.2E-3 & 1.2 & 4.4E-2 & 0.8 \\
128 & 2.0E-3 & 2.5 & 1.8E-2 & 2.4 & 1.4E-3 & 1.2 & 2.6E-2 & 0.8 \\
256 & 2.6E-4 & 2.9 & 2.5E-3 & 2.8 & 6.1E-4 & 1.2 & 1.5E-2 & 0.8 \\
\hline
\end{tabular*}
\end{center}
~

\subsection{Sensitivity to $C_{VL}$, $C_{PPM}$}
\noindent
Using a Gaussian wave as a test problem, the sensitivity in the extremum-preserving limiter was analyzed by plotting $\ln(\Delta x)$ versus $\ln(L_{\infty})$ for multiple values of $C_{VL}=C_{PPM}$ between 1.25 and 5. The extremum-preserving limiter is insensitive to the limiting constants $C_{VL}$, $C_{PPM}$, where $L_{\infty}$ changed by 4\% when comparing $C_{VL}, C_{PPM} = 1.25$ and 5.

~
\\

\pagebreak


\section{Algorithms for Nonlinear Systems of Hyperbolic Conservation Laws}
\noindent
Higher-order limiting can be extended to nonlinear systems of conservation laws. In one spatial dimension, these systems having the following form:
\begin{equation}
\partdif{U}{t} + \partdif{F}{x} = 0 , ~~ U(x,t) \in \mathbb{R}^{N} , ~~ F = F(U) , ~~ F: \mathbb{R}^{N} \rightarrow \mathbb{R}^{N} , 
\end{equation}

\noindent
where the quasilinear form is:
\begin{equation}
\partdif{U}{t} + A \partdif{U}{x} = 0 , ~~ A = A(U) = \partdif{F}{U} .
\end{equation}

\noindent
$A$ has $N$ real eigenvalues corresponding to $N$ linearly independent eigenvectors:
\begin{equation}
A r^{k} = \lambda^{k} r^{k} , ~~ l^{k} A = \lambda^{k} l^{k} ,
\end{equation}
\begin{equation}
r^{k} = r^{k}(U^{k}_{i}) , ~~ l^{k} = l^{k}(U^{k}_{i}) .
\end{equation}

\begin{figure}
\begin{center}
\includegraphics[width=6in,angle=0]{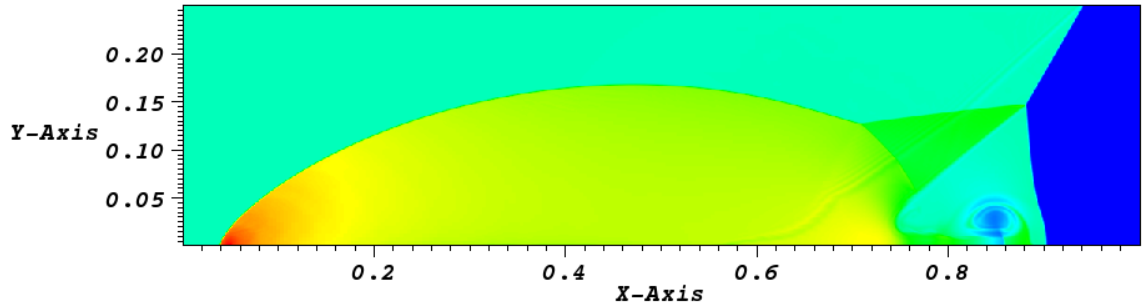}
\caption{\label{fig:chombo_full}Extremum-Preserving Limiter for the Woodward-Colella Ramp Problem. Material density for the entire domain. Computed using Chombo.}
\end{center}
\end{figure}

\begin{figure}
\begin{center}
\includegraphics[width=6in,angle=0]{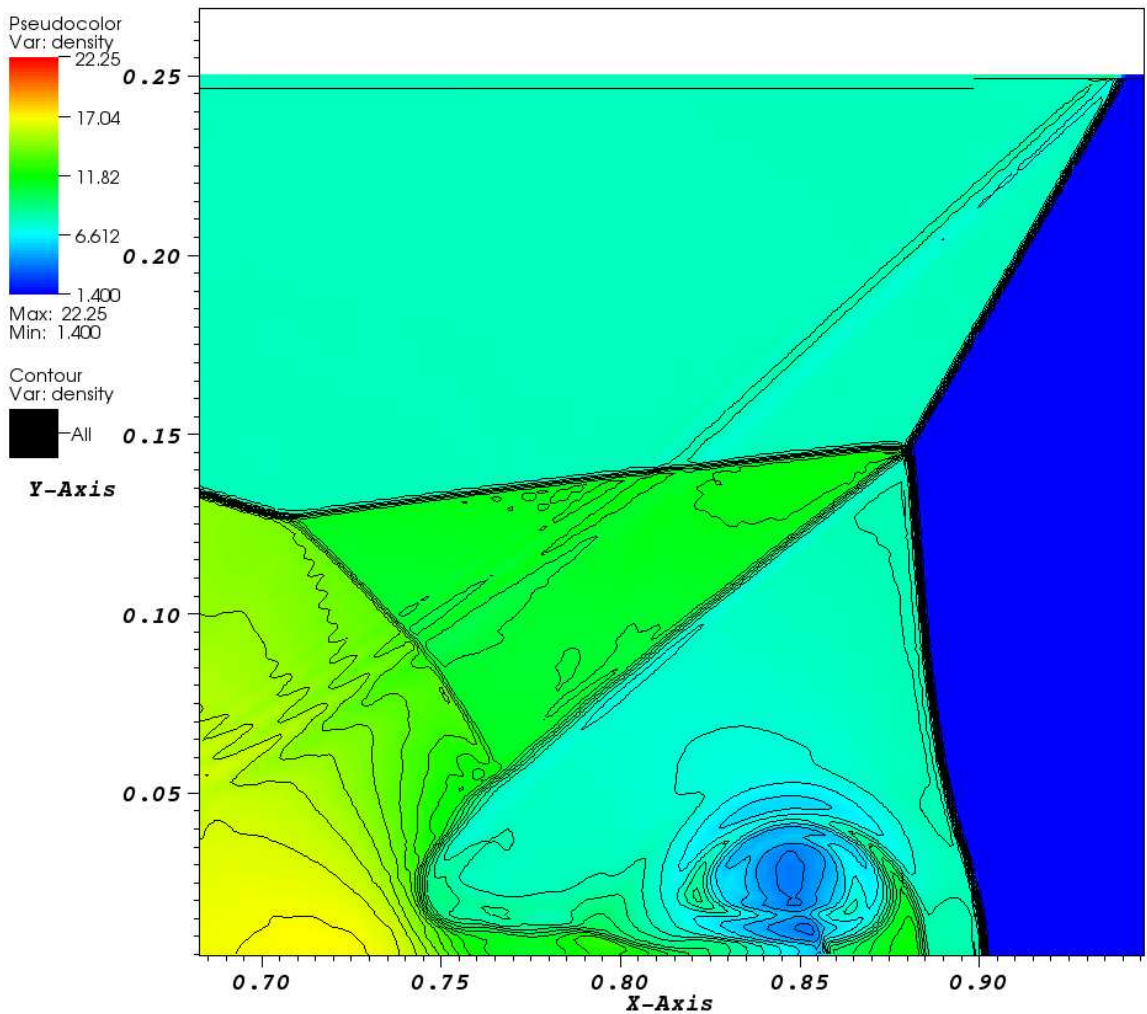}
\caption{\label{fig:chombo_zoom}Extremum-Preserving Limiter for the Woodward-Colella Ramp Problem. Material density for the double Mach region. Computed using Chombo.}
\end{center}
\end{figure}

\noindent
Like the limiting done for scalar advection, each of the variations (VL, PPM, VL+PPM) can be implemented for nonlinear systems of hyperbolic conservation laws. The limiters are either applied componentwise to the primitive variables or applied to one characteristic field at a time. The former approach was implemented in Chombo (adaptive mesh refinement infrastructure for solving a wide variety partial differential equations) \cite{chombo} and Athena (magnetohydrodynamics code for astrophysical applications) \cite{athena}. Results for extremum-preserving limiters applied to nonlinear systems of hyperbolic conservation laws are shown in \cite{athena}. In particular, \cite{athena} presents the Shu-Osher shock entropy wave interaction \cite{shuOsher:1989}, where the numerical solution obtained using an extremum-preserving limiter is comparable to WENO3 and WENO5 \cite{balsaraShu:2000}. These results are replotted below in Figures \ref{fig:ath_conv} and \ref{fig:ath_ext} with the permission of the authors. Figures \ref{fig:chombo_full} and \ref{fig:chombo_zoom} show results from the Woodward-Colella ramp problem \cite{woodwardColella:1984} that were computed by Chombo using the unsplit PPM method of Miller-Colella \cite{millerColella:2002} and the extremum-preserving limiters of Colella-Sekora \cite{colellaSekora:2007}. This calculation employed a $64 \times 16$ base grid with two levels of adaptive mesh refinement such that there is a factor of four between each level. Thus, there is an effective resolution of $1024 \times 256$. The two figures show material density for the entire domain and the double Mach region, respectively. From these plots, it is clear that extremum-preserving limiters are robust enough to handle multidimensional shocks.


\section*{Acknowledgment}
\noindent
MS acknowledges support from the DOE CSGF Program which is provided under grant DE-FG02-97ER25308.


\begin{figure}
\begin{center}
\includegraphics[width=4in,angle=0]{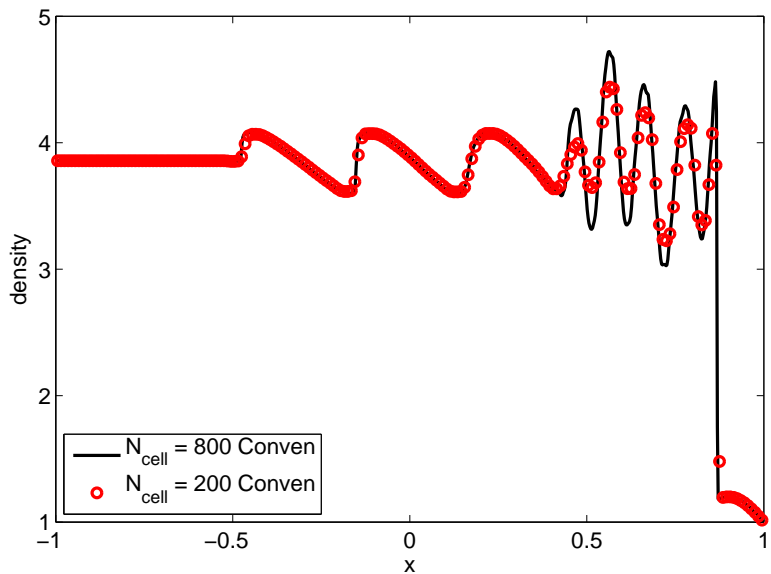}
\caption{\label{fig:ath_conv}Conventional Limiter for the Shu-Osher Shock Entropy Wave Interaction. Computed using Athena.}
\end{center}
\end{figure}

\begin{figure}
\begin{center}
\includegraphics[width=4in,angle=0]{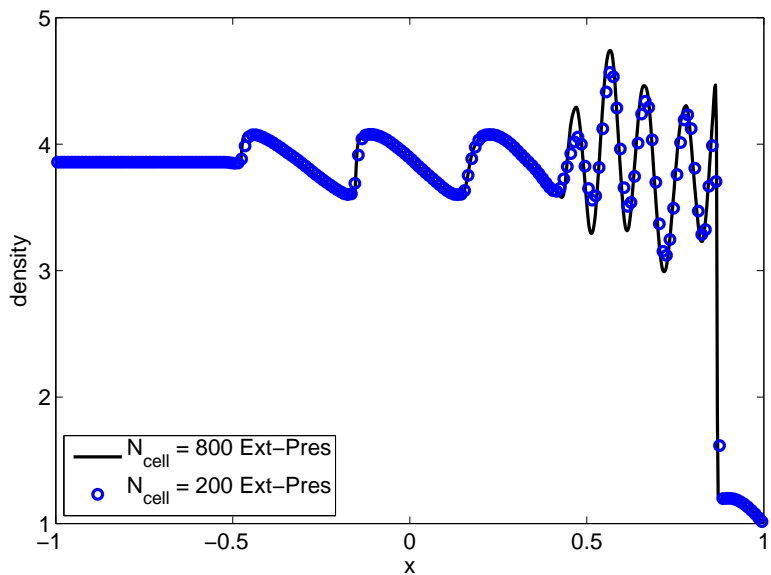}
\caption{\label{fig:ath_ext}Extremum-Preserving Limiter for the Shu-Osher Shock Entropy Wave Interaction. Computed using Athena.}
\end{center}
\end{figure}

\pagebreak




\begin{thebibliography}{12}
\bibitem {borisBook:1976} J P Boris and D L Book. Flux-corrected transport. III. Minimal-error FCT algorithms. \textit{Journal of Computational Physics}, 20:397-431, 1976. 
\bibitem {vanLeer:1977} B van Leer. Towards the ultimate conservative difference scheme. IV. A new approach to numerical convection. \textit{Journal of Computational Physics}, 23:276-299, 1977.
\bibitem {zalesak:1979} S T Zalesak. Fully multidimensional flux-corrected transport algorithms for fluids. \textit{Journal of Computational Physics}, 31:335-362, 1979.
\bibitem {harten:1982} A Harten. High resolution schemes for hyperbolic conservation laws. \textit{Journal of Computational Physics}, 49:357-393, 1983.
\bibitem {ENO} A Harten, B Engquist, S Osher, and S R Chakravarthy. Uniformly high order accurate essentially non-oscillatory schemes, III. \textit{Journal of Computational Physics}, 77:231-303, 1987.
\bibitem {WENO} G S Jiang and C W Shu. Efficient implementation of weighted ENO schemes. \textit{Journal of Computational Physics}, 126:202-228, 1996.
\bibitem {huynh} H T Huynh. Accurate upwind methods for the Euler equations. \textit{SIAM Journal on Numerical Analysis}, 32:1565-1619, 1995.
\bibitem {riderETAL:2007} W J Rider, J A Greenough, and J R Kamm. Accurate monotonicity- and extrema-preserving methods through adaptive nonlinear hybridizations. \textit{Journal of Computational Physics}, 225:1827-1848, 2007.
\bibitem {colellaWoodward:1984} P Colella and P R Woodward. The Piecewise Parabolic Method (PPM) for gas-dynamical simulations. \textit{Journal of Computational Physics}, 54:174-201, 1984.
\bibitem {millerColella:2002} G H Miller and P Colella. A conservative three-dimensional eulerian method for coupled solid-fluid shock capturing. \textit{Journal of Computational Physics}, 183:26-82, 2002.
\bibitem {colellaSekora:2007} P Colella and M D Sekora. A limiter for PPM that preserves accuracy at smooth extrema. \textit{Journal of Computational Physics}, 227:7069-7076, 2008.
\bibitem {baradColella:2006} M Barad and P Colella. A fourth-order accurate local refinement method for Poisson's equation. \textit{Journal of Computational Physics}, 209:1-18, 2005.
\bibitem {woodward:1986} P R Woodward. Piecewise-parabolic methods for astrophysical fluid dynamics. K. H. A. Winkler and M. L. Norman, editors. \textit{Astrophysical Radiation Hydrodynamics}, pg 245-326, 1986.
\bibitem {zalesak:1984} S Zalesak. R Vichnevetsky and R Stepleman, editors. \textit{Advances in Computer Methods for Partial Differential Equations V}, 1984.
\bibitem {chombo} Applied Numerical Algorithms Group, Lawrence Berkeley National Laboratory. http://seesar.lbl.gov/ANAG/chombo/ .
\bibitem {athena} J M Stone, T A Gardiner, P Teuben, J F Hawley, and J B Simon. Athena: a new code for astrophysical MHD. \textit{Astrophysical Journal Supplement Series}, 178:137-177, 2008 (pending publication).
\bibitem {shuOsher:1989} C W Shu and S Osher. Efficient implementation of essentially non-oscillatory shock-capturing schemes, II. \textit{Journal of Computational Physics}, 83:32-78, 1989.
\bibitem {balsaraShu:2000} D S Balsara and C W Shu. Monotonicity preserving weighted essentially
non-oscillatory schemes with increasingly high order of accuracy. \textit{Journal of Computational Physics}, 160:405-452, 2000.
\bibitem {woodwardColella:1984} P Woodward and P Colella. The numerical simulation of two-dimensional fluid flow with strong shocks. \textit{Journal of Computational Physics}, 54:115-173, 1984.
\end{thebibliography}
\end{document}